\documentstyle[PASJadd,epsfig]{PASJ95}
\begin{document}

\setcounter{page}{000}
\title{Is Galactic Star Formation Activity
Increased  \\
During Cluster Mergers? 
}
\author{ Yutaka {\sc Fujita},$^1$
Motokazu {\sc Takizawa},$^2$
Masahiro {\sc Nagashima},$^1$ \\
and Motohiro {\sc Enoki}$^1$
\\[12pt]
$^1${\it Department of Earth and Space Science, Graduate School of
  Science, Osaka University, }\\
  {\it Machikaneyama-cho, Toyonaka, Osaka, 560-0043, Japan}
\\
{\it  E-mail(YF): fujita@vega.ess.sci.osaka-u.ac.jp}
\\[3pt]
$^2${\it Department of Astronomy, Faculty of Science, Kyoto
University, 
Sakyo-ku, Kyoto 606-8502, Japan}} 
\abst{We have investigated the effect of pressure from intracluster
  medium (ICM) on disk galaxies in merging clusters. The ram-pressure on
  the galaxies rapidly increases when two clusters collide. This leads
  to stripping of the interstellar medium (ISM) and decrease of star
  formation rate (SFR) of the galaxies. On the other hand, the increase
  of SFR caused by compression of ISM is less significant. Thus, cluster
  merger does not trigger, but weakens star formation activity of the
  galaxies. In the central region of the colliding clusters, blue
  galaxies with high velocity should exist, although most of galaxies
  become red. Following the decrease of blue galaxy fraction in the
  clusters, the fraction of post-starburst galaxies increases. After
  merger, many galaxies in the cluster restart star formation activity
  and the segregation of blue and red galaxies becomes
  prominent.}

\kword{galaxies: clusters of --- galaxies: evolution --- galaxies:
intergalactic medium (OU-TAP 98)}

\maketitle

\section{Introduction}

According to the hierarchical clustering scenario, clusters of galaxies
are formed through subcluster mergers. The influence of cluster merger
on the intracluster medium (ICM) has been investigated in detail through
the comparison between hydrodynamic/N-body simulations and X-ray
observations. The simulations predict that the collision of large
clusters gives rise to distorted X-ray contours and high temperature gas
(e.g. Schindler, M\"{u}ller 1993; Burns et al. 1994; Ishizaka, Mineshige
1996; Roettiger et al. 1997; Ricker 1998; Takizawa 1999).  Recent X-ray
observations have confirmed that many clusters have the complex
structure predicted by the simulations (e.g. Fujita et al. 1996; Honda
et al. 1996; Knopp et al. 1996; Donnelly et al. 1998;
Markevitch et al. 1999).  On the other hand, the influence of merger on
the galaxies in the clusters is not understood. Since cluster merger
drastically changes the environment of the galaxies, especially static
and ram-pressure on the galaxies, in a short time ($\ltsim 10^9$ yr), we
can expect that it causes observable change in star formation rate (SFR)
of the galaxies. However, it is not obvious whether cluster merger
increases or decreases SFR of the galaxies as follows. Cluster merger
rapidly raises the static and ram-pressure from ICM. As a result,
interstellar medium (ISM) of the galaxies is expected to be compressed
and star formation activity may be triggered (e.g. Evrard. 1991; Wang et
al. 1997). In fact, in several merging clusters, there are galaxies
having the abnormal spectrum which reflects recent star formation
(e.g. Caldwell et al. 1993; Wang et al. 1997), although Tomita et
al. (1996) find that this is not the case for a merging cluster A168. On
the contrary, cluster merger may reduce SFR of the galaxies because
ram-pressure strips their ISM away. Thus, in order to investigate the
effect of pressure on galaxies, SFR of galaxies must be quantitatively
estimated. Using the simple model of molecular cloud evolution, Fujita,
Nagashima (1999) have quantitatively estimated the SFR of a disk galaxy
under the pressure from ICM. However they consider only a radially
infalling galaxy; they do not predict the evolutions of all galaxies in
the cluster.
 
In this letter, we investigate the evolution of SFR of disk galaxies
when two clusters collide and merge. Moreover, we calculate the color
distribution of the galaxies in the clusters.  We only consider the
effect of pressure from ICM; we do not consider the effect of tidal
force from cluster potential and other galaxies for simplicity, although
it may cause starburst (Bekki 1999). This is because it is difficult to
calculate the intermittent influence of tidal force on the internal
structures of hundreds of galaxies.

\section{Models}

We consider the merger of two typical clusters. In order to calculate
the evolution of ICM, we use the smoothed-particle hydrodynamics (SPH)
method (Monaghan 1992). We treat ions and electrons separately based on
the model of Takizawa (1999), although the two-temperature nature does
not affect the SFR of galaxies significantly. Collisionless particles
corresponding to dark matter and galaxies are also considered.  The
initial conditions for ICM and collisionless particles are the same as
those of Run B in Takizawa (1999) except for the radii of the two
clusters, $r_{\rm out}$. Since we assume that $r_{\rm out}$ is ten times
the core radius, which is two times larger than that in Takizawa (1999),
the total masses of the two clusters are also larger. Their masses are
$8\times 10^{14}$ and $2\times 10^{14}\rm\; M_{\odot}$. The gas fraction
of the clusters is 0.1, which is supported by recent observations if
$H_0\sim 75\rm\: km\; s^{-1} Mpc^{-1}$ (e.g. Ettori, Fabian 1999). The
results in the next section are not sensitive to the fraction within the
range of recent observational results. At $t=0$, the separation of the
two clusters is 3.3 Mpc. We randomly pick out 125 particles from the
collisionless particles (100 for the larger cluster and 25 for the
smaller cluster) as disk galaxies. We calculate the orbits of these
`galaxies' and the pressure from the surrounding ICM.

The effects of static and ram-pressure on the SFR of disk galaxies are
estimated by the model of Fujita (1998) and Fujita, Nagashima (1999). In
this model, the SFR is derived by calculating the evolution of each
molecular cloud using the relations for a vilialized cloud. We think
that the model is superior to the approach based on the Schmidt law
(Schmidt 1959), which has dominated in this field. This is because while
the latter gives the same SFR regardless of the pressure of ISM for a
fixed density, the former does not. Moreover the latter does
not discriminate between HI gas and molecular clouds. The model adopted
here treats them separately, although we sometimes call them together
interstellar medium (ISM). Note that the SFR derived through the model
adopted here is less sensitive to pressure variation compared to that
through the model based on Schmidt law in which the density of ISM is
assumed to be proportional to the pressure.

The condition of ram-pressure stripping is
\begin{eqnarray}
  \label{eq:strip}
  &\rho_{\rm ICM}v_{\rm rel}^2   \nonumber\\ 
  >& 2\pi G \Sigma_{\star} \Sigma_{\rm HI} \nonumber\\
  =& v_{\rm rot}^2 R^{-1} \Sigma_{\rm HI} \label{eq:grav2} \nonumber\\
  =& 2.1\times 10^{-11}{\rm dyn\: cm^{-2}}
               \left(\frac{v_{\rm rot}}{220\rm\; km\: s^{-1}}\right)^2
               \nonumber\\
   &   \times \left(\frac{R}{10\rm\; kpc}\right)^{-1}
               \left(\frac{\Sigma_{\rm HI}}{8\times 10^{20} 
                   m_{\rm H}\;\rm cm^ {-2}}\right) \label{eq:grav3}\:, 
\end{eqnarray}
where $\rho_{\rm ICM}$ is the density of ICM, $v_{\rm rel}$ is the
velocity of a galaxy relative to the surrounding ICM, $G$ is the
gravitational constant, $\Sigma_{\star}$ is the gravitational surface
mass density, $\Sigma_{\rm HI}$ is the surface density of the HI gas,
$v_{\rm rot}$ is the rotation velocity, and $R$ is the characteristic
radius of the galaxy (Gunn, Gott 1972; Fujita, Nagashima 1999). Abadi et
al. (1999) numerically confirm that this analytic relation provides a
good approximation. After this condition is satisfied, the formation of
molecular cloud is assumed to stop; the gas ejected from stars and
supplied from destroyed molecular clouds {\it directly} flows into ICM.
Note that ISM in the central region of a galaxy ($\ltsim 2$ kpc) is not
stripped because of large gravitational force. However, the mass is
generally far smaller than the total mass of ISM (e.g. Struck-Marcell
1991). Thus, we ignore its contribution to the star formation activity
of the galaxy.

If a stripped galaxy reaches the outer part of the cluster, the galaxy
may recover ISM. Since Fujita, Nagashima (1999) do not take account of
recovery of ISM, we adopt the condition of recovery,
%%%% Eq.2
\begin{eqnarray}
  \label{eq:ret}
   &\rho_{\rm ICM}v_{\rm rel}^2  \nonumber\\
  < & \frac{16}{v_{\rm rel}}  \frac{S}{\pi R^2}v_{\rm rot}^2 \nonumber\\
  = & 2.0\times 10^{-11} {\rm dyn\; cm^{-2}}
    \left(\frac{v_{\rm vel}}{500\rm\; km\; s^{-1}}\right)^{-1} 
       \nonumber\\
    & \times
    \left(\frac{S}{6\rm\; M_{\odot}\;yr^{-1}}\right)
    \left(\frac{R}{10\rm\; kpc}\right)^{-2}
    \left(\frac{v_{\rm rot}}{220\rm\; km\; s^{-1}}\right)^{2}
\:,
\end{eqnarray}
where $S$ is the gas supply from stars and destroyed molecular clouds
(Takeda et al. 1984). Although this analytic expression assumes
spherical symmetry and may not be exact in the case of a disk galaxy,
the following results do not significantly alter even if we change the
coefficient of the right hand of relation (\ref{eq:ret}) by a factor of
five. After the galaxy satisfies this condition, gas ejected from stars
is trapped in the potential well of the galaxy, and molecular cloud
formation is resumed.

We start to calculate the SFR of the model galaxies at $t=1$ Gyr. The
initial mass of molecular gas and HI column density are $2.5\times
10^9\rm\; M_{\odot}$ and $8\times 10^{20} m_{\rm H}\;\rm cm^ {-2}$,
respectively. Moreover, we take $S=6\rm\; M_{\odot}\:yr^{-1}$, $R=10$
kpc, and $v_{\rm rot}=220\rm\;km\; s^{-1}$. Although the parameters are
the typical ones for our Galaxy (e.g. Binney, Tremaine 1987), the
following results do not change significantly even if we take the ones
for typical galaxies whose masses are five times smaller. We have
confirmed that the evolutions of the SFR and ISM are not sensitive to
the initial time of the calculation $\gtsim 1$ Gyr after the calculation
starts.  Using the obtained SFR and the population synthesis code made
by Kodama, Arimoto (1997), we also investigate the evolution of color of
galaxies.

\section{Results}
\label{sec:results}

Figure 1a shows the X-ray contours and positions of galaxies
at $t=3.6$ Gyr, where the two clusters have just passed each other. The
origin of the figure is the center of gravity of the clusters.
%%%% Fig.1a, 1b
We define a `post-starburst galaxy' (PSB) as the galaxy whose SFR
reduces to less than $1/3$ of that for $10^8 - 10^9$ yr before the
observation time.  Figure~1a shows that red ($B-V>0.7$) galaxies and
PSBs are concentrated in the central region of the cluster. These
features are always seen during merger. Although several blue
($B-V<0.7$) galaxies are also seen in the central region, they do not
gather at a specific position. We present the velocity distributions of
blue and red galaxies within $0.5$ Mpc from the center of the merged
cluster at $t=3.6$ Gyr in figure~2.
%%%% Fig.2
Since the average velocity of
blue galaxies in this region is $\sim 2500\rm\; km\; s^{-1}$, they pass
each other simultaneously. Note that in figure 2 the average velocity of
red galaxies is $\sim 1800\rm\; km\; s^{-1}$. This reflects that the
blue galaxies at the central region of the cluster are the ones that
can reach the cluster center before they become red although their ISM
is stripped. In figure 1b, we present the state of the cluster after it
is nearly relaxed ($t=5$ Gyr). Segregation of blue and red galaxies is
noticeable.  This is quantitatively shown in figure~3. 
%%%% Fig.3
The number of 
blue galaxies in the central region of cluster ($\ltsim 0.7$ Mpc) at
$t=5$ Gyr is smaller than that at $t=3.6$ Gyr.  This is
because there are few galaxies rapidly infalling into the center of the
merged cluster at $t=5$ Gyr.

To see the evolution of galaxies in detail, we show the fraction of blue
galaxies and PSBs in figure~4. The evolution of the latter is
calculated only for $t>3$ Gyr to avoid the influence of initial
conditions. The median static and ram-pressures on galaxies are
shown for comparison. They temporarily increase at $t\sim 3.6$ Gyr
when two clusters collide. At that time, the HI gas of most galaxies is
stripped because of the increase of ram-pressure.  After that, new
molecular clouds are not produced; the existing ones disappear within
$\sim 10^8$ yr because of consumption by star formation and destruction
by young stars (see Fujita, Nagashima 1999).  Since molecular clouds are
used to make stars, the SFR of galaxies and fraction of blue galaxy
decrease as molecular clouds disappear. Although the static and
ram-pressure compress ISM of galaxies and trigger the star formation
activity before the stripping occurs, the duration of activity is short
($\ltsim 0.4$ Gyr); the activity does not affect the color distribution
of galaxies in clusters significantly. Thus, cluster merger does not
trigger, but weakens star formation activity of the galaxies. At $t\sim
4$ Gyr, the two clusters once come apart. The ICM of the two clusters
expands and their relative velocity reduces. Since the average
ram-pressure significantly decreases, the ISM of galaxies recovers and
star formation restarts. Thus, the fraction of blue galaxies returns to
the initial value ($\sim 60$\%). Note that the fraction of blue galaxies
is over 40\% even when clusters are colliding (figure~1). This means
that cluster merger does not significantly affect galaxies in the outer
region of the clusters.

Figure~4 clearly shows that the fraction of PSBs increases from 30\% to
60\%, immediately following the merger. These PSBs are the galaxies
whose SFR decreases because of ram-pressure stripping. In fact, the
fraction of PSBs begins to increase after that of blue galaxies
decreases. However, it may not be easy to use PSBs as the probe of
cluster merger, because they always exist in clusters. This reflects
that some of cluster galaxies have radial orbits and fall into the
center of the cluster regardless of cluster merger. Thus, in order to
know the relation between PSBs and cluster merger observationally, it is
required to compare the PSB fractions of merging clusters with those of
non-merging clusters statistically.

\section{Summary}
\label{sec:summary}

We have investigated the effect of pressure on the galaxies when two
clusters merge. We find that because of ram-pressure stripping, star
formation rate of most of galaxy decreases during merger contrary to the
speculation of Evrard (1991) and Wang et al. (1997). Some blue galaxies
can reach the central region of the merging clusters before they become
red because of their high velocities. By observing velocities, these
galaxies would be distinguished from the blue galaxies in which star
formation is triggered by tidal force from the gravitational field of
the cluster because the tidal interaction is effective when a galaxy
moves slowly. Following the decrease of blue galaxy fraction of the
clusters, the fraction of post-starburst galaxies increases. After the two
clusters pass by, star formation restarts in the galaxies because the
ram-pressures decrease. When a quasi-equilibrium state is reached, the
segregation of blue and red galaxies becomes prominent.

\par
\vspace{1pc} \par
This work was supported in part by the JSPS Research Fellowship for
Young Scientists.

\section*{References}
\small

\re Abadi M.G., Moore B., Bower R.G.\ 1999, MNRAS in press 
(astro-ph/9903436)

\re Bekki K.\ 1999, ApJL 510, L15

\re Binney, J., Tremaine, S. 1987, Galactic Dynamics

\re Burns J.O., Roettiger K., Ledlow M., Klypin A.\ 1994, ApJL 427,
L87

\re Caldwell H., Rose J.A., Sharples R.M., Ellis R.S., Bower R.G.\ 
1993, AJ 106, 473

\re Donnelly R.H., Markevitch M., Forman W., Jones C., David L.P.,
Churazov E., Gilfanov M.\ 1998, ApJ 500, 138

\re Ettori S., Fabian A.C.\ 1999, MNRAS in press (astro-ph/9901304)

\re Evrard A.E.\ 1991, MNRAS 248, 8p

\re Fujita Y.\ 1998, ApJ 509, 587

\re Fujita Y., Koyama K., Tsuru T., Matsumoto H.\ 1996, PASJ 48, 191

\re Fujita Y., Nagashima M., 1999, ApJ in press (astro-ph/9812378)

\re Gunn J.E., Gott J.R. 1972, ApJ 176, 1

\re Honda H., Hirayama M., Watanabe M., Kunieda H., Tawara Y.,
Yamashita K., Ohashi T., Hughs, J.P., Henry, J.P.\ 1996, ApJL 473,
L71

\re Ishizaka C., Mineshige S.\ 1996, PASJ 48, L37

\re Knopp G.P., Henry J.P., Briel U.G.\ 1996, ApJ 472, 125

\re Kodama T., Arimoto N.\ 1997, A\&A 320, 41

\re Markevitch M., Sarazin C.L., Vikhlinin A.\ 1999, ApJ, in press
(astro-ph/9812005)

\re Monaghan J.J.\ 1992, ARA\&A 30, 543

\re Ricker P.M.\ 1998, ApJ 496, 670

\re Roettiger K., Loken C., Burns J.O.\ 1997, ApJS 109, 307

\re Schindler S., M\"{u}ller E.\ 1993, A\&A 272, 137

\re Schmidt M.\ 1959, ApJ 344, 685

\re Struck-Marcell 1991, ApJ, 368, 348

\re Takeda H., Nulsen P.E.J., Fabian A.C.\ 1984, MNRAS 208, 261

\re Takizawa M.\ 1999, ApJ in press (astro-ph/9901314)

\re Tomita A., Nakamura F.E., Takata T., Nakanishi K., Takeuchi T.,
Ohta K., Yamada T.\ 1996, AJ 111, 42

\re Wang Q.D., Ulmer M.P., Lavery R.J.\ 1997, MNRAS 288, 702

\label{last}

\newpage

\begin{figure}
\centering \epsfig{figure=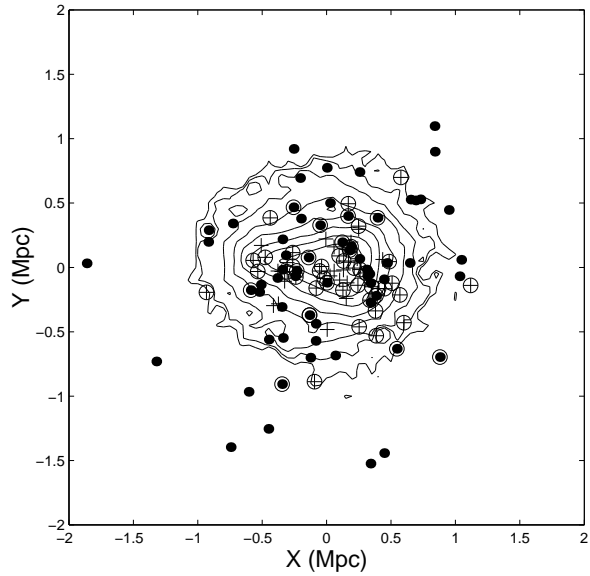, width=8cm} 
\caption{(a)
The X-ray surface
brightness and the positions of galaxies at $t=3.6$ Gyr. The crosses and
filled circles indicate red ($B-V>0.7$) and blue ($B-V<0.7$) galaxies,
respectively. Open circles indicate post-starburst galaxies.}
\end{figure}

\begin{figure}
\centering \epsfig{figure=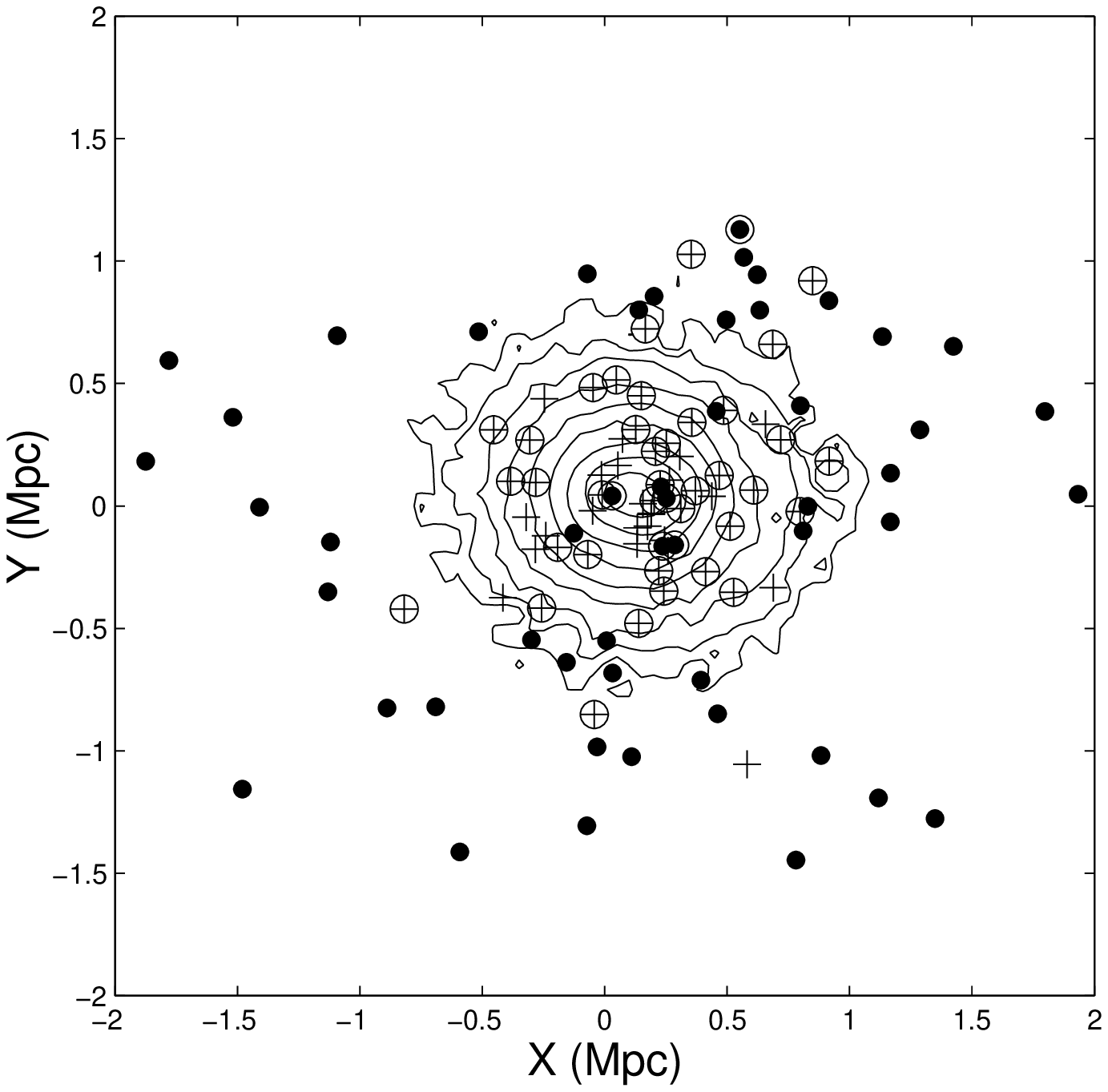, width=8cm} 

\vspace{3mm}

\footnotesize{Fig. 1.. (b)
Same as in figure 1 but for $t=5$ Gyr.}
\end{figure}

\begin{figure}
\centering \epsfig{figure=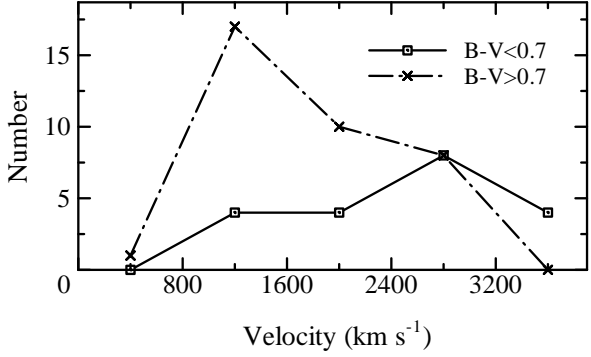, width=8cm} \caption{
Histogram showing the distribution of velocity relative to the space
 coordinate at $t=3.6$ Gyr. The galaxies within 0.5 Mpc from the
 origin are picked out.
}
\end{figure}

\begin{figure}
\centering \epsfig{figure=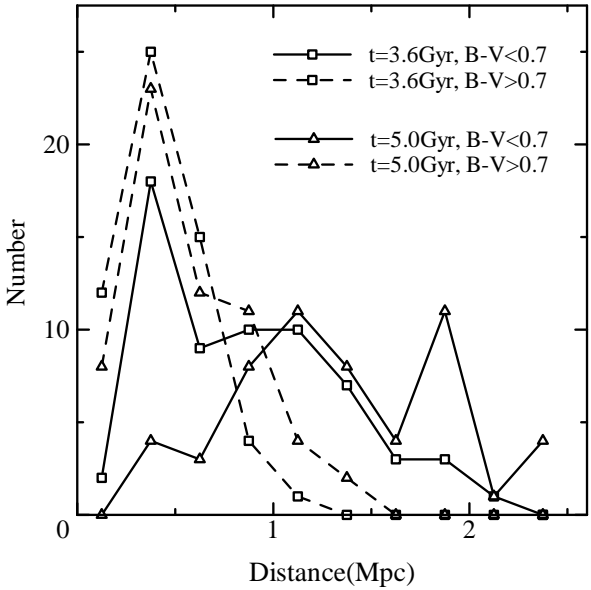, width=8cm} \caption{
Histogram showing the distribution of distance from the origin.}
\end{figure}

\begin{figure}[t]
\centering \epsfig{figure=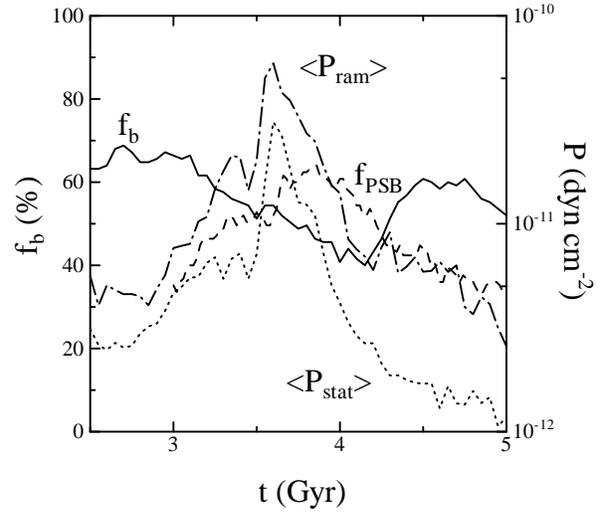, width=8cm} \caption{
The fraction of blue galaxies $f_b$ and post-starburst
    galaxies $f_{\rm PSB}$. The median static pressure $\langle
    P_{\rm stat} \rangle$ and ram-pressure $\langle P_{\rm ram}
    \rangle$ are also presented.}
\end{figure}

\end{document}